\documentclass[twocolumn,showpacs,prl,superscriptaddress,floatfix]{revtex4}

\usepackage{color}
\usepackage{tabularx}
\usepackage{epsfig}
\usepackage{amsmath}
\usepackage{amssymb}
\usepackage{graphicx}
\usepackage{wasysym}
\usepackage{dcolumn}
\usepackage{bm}
\usepackage{subfigure}
\usepackage{psfrag}
\usepackage{subfigure}
\usepackage{times}

\newcommand{\pq}{P(q)}
\newcommand{\qea}{q_{\rm EA}}

\begin{document}

\title{Evidence of non-mean-field-like low-temperature behavior in
the Edwards-Anderson spin-glass model}

\author{B.~Yucesoy} 
\affiliation{Physics Department, University of Massachusetts, Amherst, 
MA 01003 USA}

\author{Helmut G.~Katzgraber}
\affiliation {Department of Physics and Astronomy, Texas A\&M University,
College Station, Texas 77843-4242, USA}
\affiliation {Theoretische Physik, ETH Zurich, CH-8093 Zurich, Switzerland}

\author{J.~Machta}
\affiliation{Physics Department, University of Massachusetts, Amherst,
MA 01003 USA}

\date{\today}

\begin{abstract}

The three-dimensional Edwards-Anderson and mean-field
Sherrington-Kirkpatrick Ising spin glasses are studied via large-scale
Monte Carlo simulations at low temperatures, deep within the spin-glass
phase. Performing a careful statistical analysis of several thousand
independent disorder realizations and using an observable that detects
peaks in the overlap distribution, we show that the
Sherrington-Kirkpatrick and Edwards-Anderson models have a distinctly
different low-temperature behavior. The structure of the spin-glass
overlap distribution for the Edwards-Anderson model suggests that its
low-temperature phase has only a single pair of pure states.

\end{abstract}


\maketitle

Spin glasses \cite{binder:86} have been the subject of intense study and
controversy for decades.  These models are perhaps the simplest,
physically-motivated examples of frustrated systems in classical
statistical mechanics. Given their wide applicability across
disciplines, it is important that their behavior is understood.  Despite
four decades of research, the low-temperature phase of short-range spin
glasses is poorly understood.  Here we study both the three-dimensional
(3D) Edwards-Anderson (EA) Ising spin glass \cite{edwards:75} and the
Ising spin glass on a complete graph---known as the
Sherrington-Kirkpatrick (SK) model \cite{sherrington:75}---in an effort
to gain a deeper understanding of the low-temperature spin-glass state.
Our results suggest that these models are qualitatively different at low
temperatures.

Parisi's solution of the SK model \cite{parisi:79,parisi:83} involves an
unusual form of symmetry breaking among replicas. These were originally
introduced to carry out the disorder average of the logarithm of the
partition function. The low-temperature phase of the model within the
replica symmetry breaking (RSB) solution \cite{parisi:79,parisi:83} has
several unusual features such as the breakdown of self-averaging and the
co-existence of a countable infinity of pure states in the thermodynamic
limit.

\begin{figure}[!h]
\center

\subfigure{\label{subfig:P1a}\includegraphics[width=0.32\columnwidth]{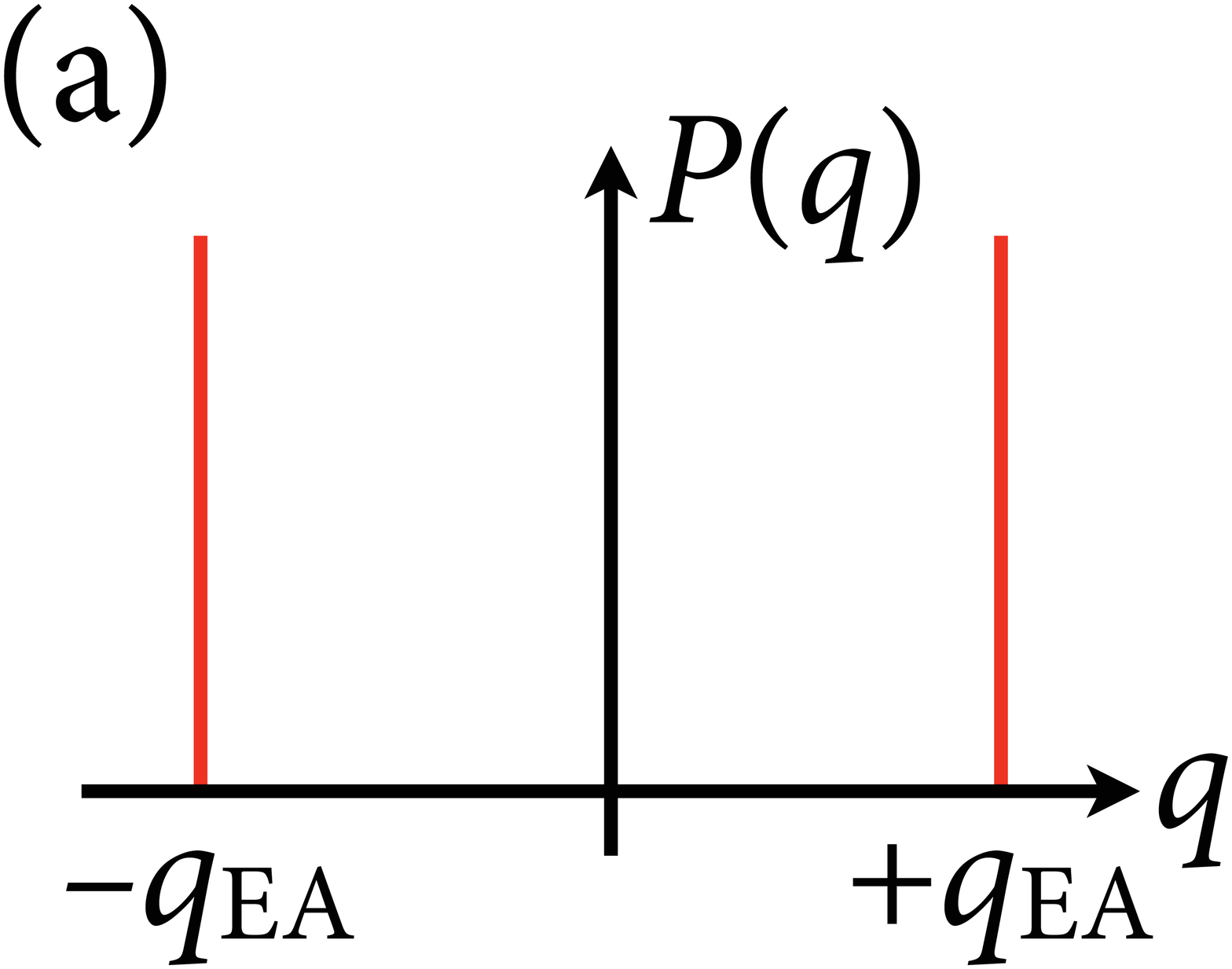}}
\subfigure{\label{subfig:P1b}\includegraphics[width=0.32\columnwidth]{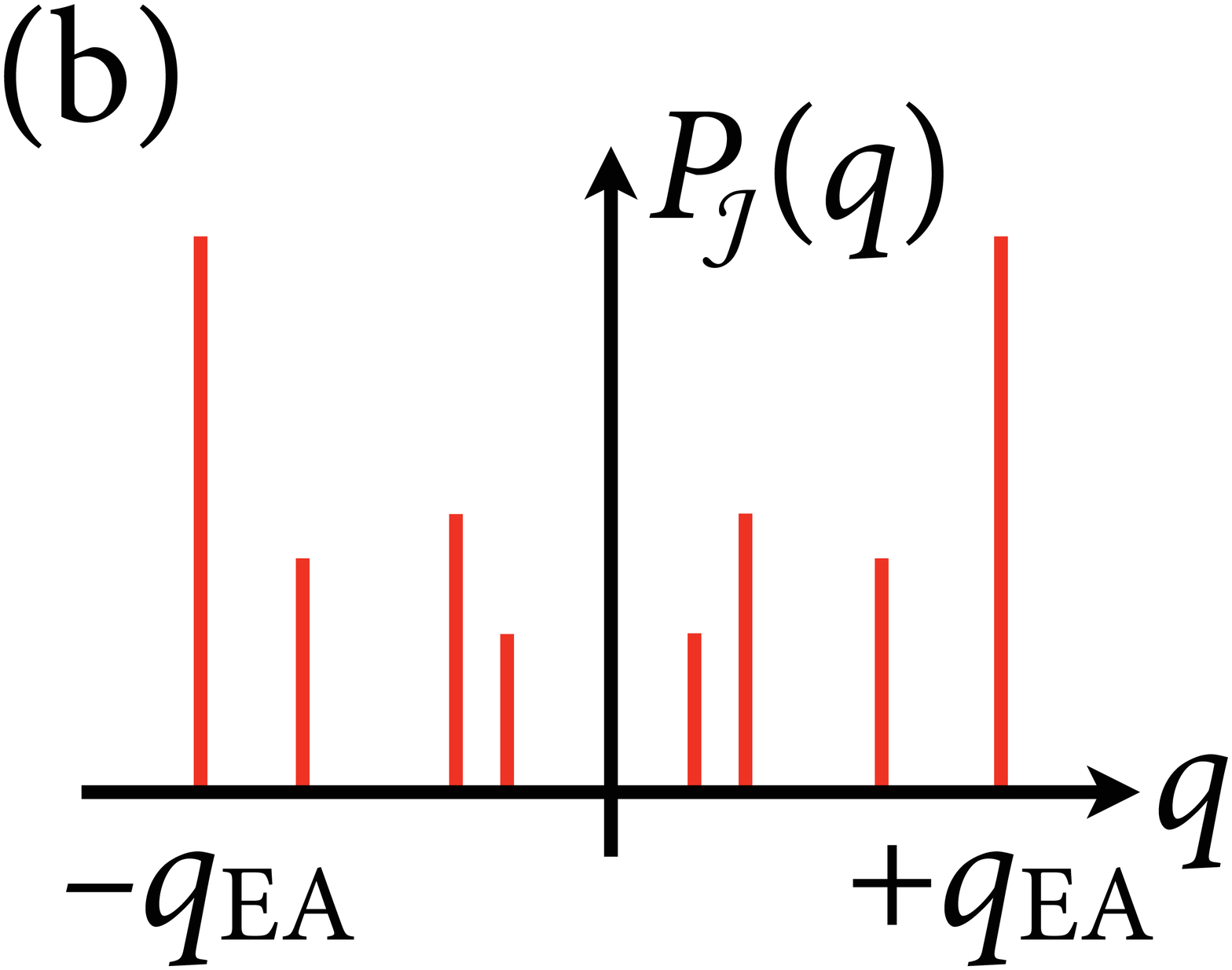}}
\subfigure{\label{subfig:P1c}\includegraphics[width=0.32\columnwidth]{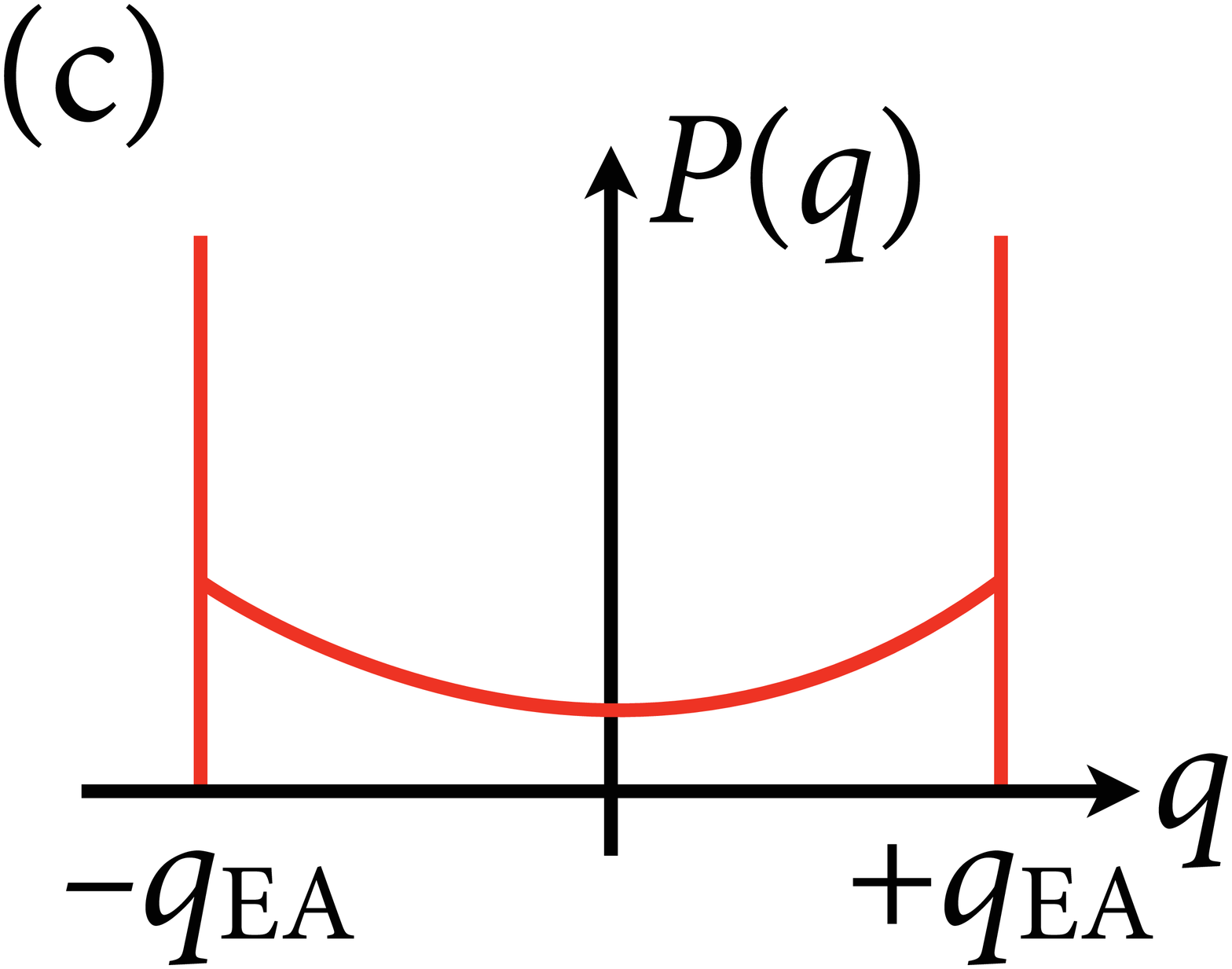}}

\vspace*{-1.5em}

\caption{(Color online)
(a) In the droplet picture $P(q)$ is trivial with one pair of pure
states. (b) In the RSB picture individual samples have many pairs of
pure states ($\delta$ functions in $P_{\cal J}(q)$).  (c) In the
RSB picture $P(q)$ is nontrivial (continuous support for $|q| < \qea$).
\label{fig:pqsketch}}
\end{figure}

There is no analytic theory for the EA model but it is well-accepted on
the basis of numerical simulations \cite{katzgraber:06} that the EA
model undergoes a continuous phase transition.  However, the
low-temperature broken-symmetry phase is not understood, even
qualitatively.  Different mutually-exclusive scenarios have been
proposed:  The replica symmetry breaking (RSB) picture is based on an
analogy with the solution of the SK model. It assumes that
self-averaging breaks down and that there are a countable infinity of
pure states in the thermodynamic limit.  A qualitatively different and
simpler picture was proposed to describe the EA model by McMillan,
Fisher and Huse, as well as Bray and Moore
\cite{mcmillan:84b,fisher:86,fisher:87,fisher:88,bray:86}.  In the
``droplet scaling'' picture the low-temperature phase is described by
one pair of pure states related by a spin flip with low-lying
excitations that are isolated, compact droplets of the opposite phase.
A central difference between the RSB and droplet pictures for the EA model
is whether there is a single pair of pure states or many pairs of pure
states for large systems, see Fig.~\ref{fig:pqsketch}.

Newman and Stein \cite{newman:92,newman:96,newman:98} explained that the
usual way of constructing the thermodynamic limit cannot be applied to
finite-dimensional spin glasses because of the possibility of a chaotic
system-size dependence in which different thermodynamic states may
appear for different system sizes. They showed that the key ideas of
RSB---non-self-averaging and a countable infinity of pure
states---cannot hold for the EA model within the na{\"{i}}ve way that
they were first proposed. However, their results do not completely rule
out a nonstandard interpretation of RSB.  They also proposed a more
plausible many-states ``chaotic pairs'' picture in which for a fixed
choice of couplings, there are many pure states but that in a single
finite volume only one pair is manifest.

\begin{figure}
\center
\includegraphics[width=0.85\columnwidth]{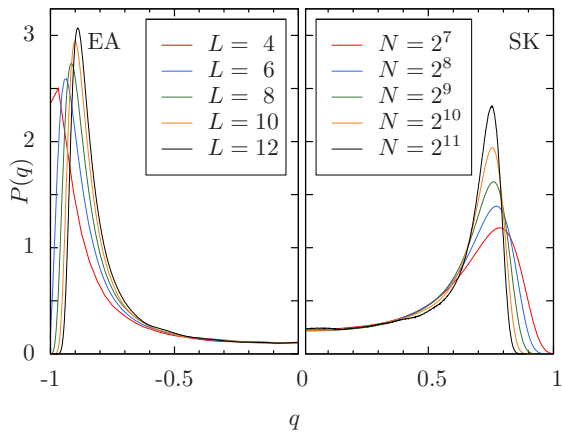}
\vspace*{-1.0em}
\caption{(Color online)
Disorder-averaged overlap probability distribution $P(q)$ for different
system sizes at $T=0.42$ and $T=0.4231$ for the EA model (left) and SK
model (right), respectively.
\label{fig:average} } 
\end{figure}

Here we report the results of large-scale Monte Carlo simulations of
both the SK and EA models. Our objective is to shed light on the
qualitative nature of the low-temperature phase of the EA model by
comparing and contrasting to the SK model.  Previous numerical studies,
e.g., \cite{katzgraber:01} using the average spin overlap
distribution suggested that both the SK and EA models are well described
by the RSB picture.  However, for the numerically-accessible system
sizes the two main peaks are still converging to $\pm \qea$ (see
Fig.~\ref{fig:average}) and therefore results might be plagued by
finite-size effects.  On the other hand, studies of the link overlap
\cite{katzgraber:01} distribution suggest agreement with the droplet
picture.  The ``trivial nontrivial" scenario
\cite{palassini:99,krzakala:00,katzgraber:01} reconciles these
numerical results by postulating that excitations are compact, as in the
droplet picture, but their energy cost is independent of system size, as
in the RSB picture. In an effort to resolve these discrepancies, here we
introduce a statistic obtained from the spin overlap distribution that
detects sharp peaks in {\em individual} samples, inspired by a recent
study on the SK model \cite{aspelmeier:08}. This statistic clearly
differentiates the RSB and droplet pictures: it converges to zero in the
large-volume limit if there is a single pair of pure states and to unity
if there are countably many. Our results for this quantity shows clear
differences between the EA and SK models.

\paragraph*{Models and Numerical Details.---}
\label{sec:model}

The SK and EA models are defined by the Hamiltonian ${\mathcal H} = -
\sum_{i,j = 1}^N J_{ij} S_i S_j$, with $S_i \in \{\pm 1\}$ Ising spins.
For the EA model the sum is over nearest neighbors on a cubic lattice of
size $N = L^3$ with periodic boundaries. The couplings $J_{ij}$ are
chosen from a Gaussian distribution with zero mean and variance unity. A
set of couplings ${\cal J} = \{J_{ij}\}$ defines a disorder realization
or, simply, a ``sample.''  For the SK model the sum is over all pairs of
spins and the $J_{ij}$ are chosen from a Gaussian distribution with zero
mean and variance $1/(N-1)$.

Ordering in spin glasses is detected from the spin overlap $q=(1/N)
\sum_i S^{\alpha}_i S^{\beta}_i$, where ``$\alpha$'' and ``$\beta$''
indicate independent spin configurations for the same sample ${\cal J}$.
The primary observable we consider for fixed ${\cal J}$ and $N$ is the
overlap probability density, $P_{\cal J}(q)$. In the high-temperature
phase there is a well-defined thermodynamic limit and $P_{\cal J}(q) \to
\delta(q)$ for $N \to \infty$ for almost every ${\cal J}$. The behavior
of $P_{\cal J}(q)$ for large $N$ and $T < T_c$, $T_c$ the critical
temperature, distinguishes the RSB picture from other theories.  If
there is only a single pair of states for each system size, $P_{\cal
J}(q)$ consists for large $N$ of a symmetric pair of $\delta$ functions
at the Edwards-Anderson order parameter $q=\pm \qea$, see
Fig.~\ref{subfig:P1a}. In the RSB picture there are many sharp peaks
symmetrically distributed in the range $-\qea < q < \qea$ as shown in
Fig.~\ref{subfig:P1b}, corresponding to multiple pairs of pure states.
In the RSB picture, the distribution of peaks depends on ${\cal J}$ but
the disorder averaged overlap distribution $P(q)$ exists, and for large
$N$ is expected take the form shown in Fig.~\ref{subfig:P1c}.

We have carried out replica exchange Monte Carlo \cite{hukushima:96}
simulations of both models. Parameters are shown in Tables
\ref{tab:paramsea} and \ref{tab:paramssk}.  For each sample we
equilibrate two independent sets of replicas to compute the overlap
distribution. Equilibration is tested for the EA and SK models using the
methods of Refs.\cite{katzgraber:01} and \cite{hukushima:98},
respectively.  The number of equilibration and data collection sweeps
are chosen to be long enough to ensure that samples are well
equilibrated and that $P_{\cal J}(q)$ is accurately measured for each
sample.  We report results for $T=0.42$ [$T=0.4231$] for the EA [SK]
model.  For the EA model, $T_c \approx 0.96$ \cite{katzgraber:06},
while for the SK model $T_c = 1$, so our simulations are at $\sim 0.4
T_c$, i.e., deep within the spin-glass phase \cite{comment:temps} where
critical fluctuations are unimportant.

\begin{table}
\caption{
EA model simulation parameters. For each number of spins $N =
L^3$ we equilibrate and measure for $2^b$ Monte Carlo sweeps.  $T_{\rm
min}$ [$T_{\rm max}$] is the lowest [highest] temperature and $N_T$
is the number of temperatures.  $N_{\rm sa}$ is the number of disorder
samples.
\label{tab:paramsea}}
{\scriptsize
\begin{tabular*}{\columnwidth}{@{\extracolsep{\fill}} r r c c c c c}
\hline
\hline
$N$ &  $L$ & $b$  & $T_{\rm min}$ & $T_{\rm max}$ & $N_{T}$ & $N_{\rm sa}$ \\
\hline
  $64$ &  $4$ & $18$ & $0.2000$       & $2.0000$        & $16$    & $4891$ \\
 $216$ &  $6$ & $24$ & $0.2000$       & $2.0000$        & $16$    & $4961$ \\
 $512$ &  $8$ & $27$ & $0.2000$       & $2.0000$        & $16$    & $5130$ \\
$1000$ & $10$ & $27$ & $0.2000$       & $2.0000$        & $16$    & $5027$ \\
$1728$ & $12$ & $25$ & $0.4200$       & $1.8000$        & $26$    & $3257$ \\
\hline
\hline
\end{tabular*}
}
\end{table}
\begin{table}
\caption{
Simulation parameters for the SK spin glass. See the Table
\ref{tab:paramsea} for details.
\label{tab:paramssk}}
{\scriptsize
\begin{tabular*}{\columnwidth}{@{\extracolsep{\fill}} r c c c c c}
\hline
\hline
   $N$ & $b$  & $T_{\rm min}$  & $T_{\rm max}$   & $N_{T}$ & $N_{\rm sa}$ \\
\hline
  $64$ & $22$ & $0.2000$       & $1.5000$        & $48$    & $5068$       \\
 $128$ & $22$ & $0.2000$       & $1.5000$        & $48$    & $5302$       \\
 $256$ & $22$ & $0.2000$       & $1.5000$        & $48$    & $5085$       \\
 $512$ & $18$ & $0.2000$       & $1.5000$        & $48$    & $4989$       \\
$1024$ & $18$ & $0.2000$       & $1.5000$        & $48$    & $3054$       \\
$2048$ & $16$ & $0.4231$       & $1.5000$        & $34$    & $3020$       \\

\hline
\hline
\end{tabular*}
}
\end{table}

\paragraph*{Results.---}
\label{sec:results}

\begin{figure}
\center
\includegraphics[width=\columnwidth]{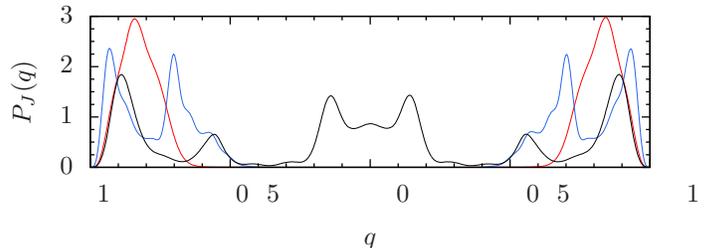}
\vspace*{-2.5em}
\caption{(Color online) 
Typical overlap distributions $P_{\cal J}(q)$ for three disorder
realizations for the EA model with $N = 8^3$ and $T = 0.42$.
\label{fig:OverlapDist}}
\end{figure}

\begin{figure}
\center
\includegraphics[width=0.85\columnwidth]{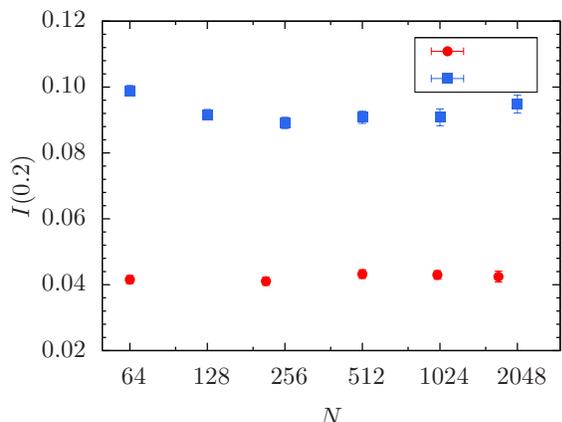}
\vspace*{-1.5em}
\caption{(Color online)
Disorder average of the weight of the overlap distribution $I(0.2)$ as a
function of $N$ for $T \approx 0.4T_c$ for both the EA and SK models.
\label{fig:p0}}
\end{figure}

Figure \ref{fig:OverlapDist} shows $P_{\cal J}(q)$ for three different
EA samples ($N = 512 = 8^3$, $T = 0.42$).  Note that $P_{\cal J}(q)$
varies considerably between samples.  Qualitatively similar overlap
distributions are seen for the SK model.  Figure \ref{fig:average}, left
panel [right panel], shows the disorder averaged overlap distribution
$\pq$ for the EA [SK] model for different system sizes at $T = 0.42$
[$T=0.4231$] \cite{comment:pq}.  At this low temperature, $\pq$
consists of large peaks at the finite-size value of the EA order
parameter, $\pm \qea(N)$.  $\pq$ is reasonably flat, non-zero, and
nearly independent of $N$ in the approximate range $-0.4\lesssim q
\lesssim 0.4$  for the sizes studied here.  We can quantify this
observation by considering the integrated overlap, $I(q_0) = \int_{|q| <
q_0} P(q) dq$.  Figure \ref{fig:p0} shows $I(0.2)$ as a function of $N$
for both the EA and SK models at $T \approx 0.4T_c$
\cite{comment:temps}. Note that $I(0.2)$ is nearly independent of $N$.
We found qualitatively similar results for other values of $q_0$ up to
$q_0 \approx 0.5$ and temperatures down to $0.2T_c$ for smaller systems.
The constancy of $I(0.2)$ has been observed in a number of studies (see
Refs.~\cite{katzgraber:01} and \cite{alvarez:10}) and is among the
strongest evidence in favor of the validity of the RSB picture for
short-range systems.

Although $I(q_0)$ in Fig.~\ref{fig:p0} is nearly constant over the range
of sizes simulated in this and other studies of the EA model, it is also
clear that, for these same sizes there are strong finite-size effects.
These corrections can be seen by looking at the size dependence of
$\qea(N)$.  The peak moves to smaller values of $\qea$ as $N$ increases,
similar to recent results \cite{alvarez:10} for larger $N$.  The
presence of these strong finite-size corrections makes the absence of
any significant $N$ dependence of $\pq$ for small $q$ surprising. In the
droplet picture, $I(q_0)$ is expected to decay with a small power of
$L$, $I(q_0) \sim TL^{-\theta}$ ($\theta \approx 0.2$ in 3D
\cite{hartmann:99}) and this slow asymptotic behavior may not set in
until large sizes.  Thus the behavior of $I(q_0)$ shown in
Fig.~\ref{fig:p0} may not be a sensitive indicator of the nature of the
low-temperature phase for system sizes currently accessible to
simulation.

To better understand the size dependence of the overlap distributions,
we go beyond disorder averages and consider other statistics obtained
from $P_{\cal J}(q)$. In particular, we identify the emergence, or not,
of $\delta$ functions in the range $-\qea<q<\qea$ as $N$ increases,
which would signal more than one pair of pure states. A finite-size
broadened $\delta$ function at $q$ is characterized by a large value of
$P_{\cal J}(q)$. To detect $\delta$-function-like behavior for finite
$N$ we consider the statistic
\begin{equation}
\Delta(q_0,\kappa) = 
{\rm Prob}\left[ \max_{|q|<q_0}
    \bigg\{
	\frac{1}{2}\big(P_{\cal J}(q)+ P_{\cal J}(-q)\big)
    \bigg\} >
\kappa \right].
\label{eq:ff}
\end{equation}
The probability is defined with respect to ${\cal J}$ and
$\Delta(q_0,\kappa)$ is the fraction of samples with at least one peak
greater than $\kappa$ in $P_{\cal J}(q)$ in the range $|q| <q_0$.
$\kappa$ is chosen to be large enough to exclude some but not all
samples.  We refer to samples counted in $\Delta(q_0,\kappa)$ as ``{\em
peaked}.''  For example, with $\kappa=1$ the sample with the central
peaks (black line) in Fig.~\ref{fig:OverlapDist} is peaked for
$q_0\gtrsim 0.1$, whereas the two other samples are not for $q_0\lesssim
0.5$.

The droplet and RSB pictures make dramatically different predictions for
$\Delta(q_0,\kappa)$. For the droplet or chaotic pairs picture there is
only a single pair of states for any large volume so that
$\Delta(q_0,\kappa) \to 0$ for any $\kappa>0$ and any $q_0 < \qea$ when
$N \to \infty$. However, for the RSB picture one expects $\delta$
functions in $P_{\cal J}(q)$ for any range of $q$, i.e.,
$\Delta(q_0,\kappa) \to 1$ as $N \to \infty$ for any $q_0$ and $\kappa >
0$.

\begin{figure}
\center
\includegraphics[width=0.85\columnwidth]{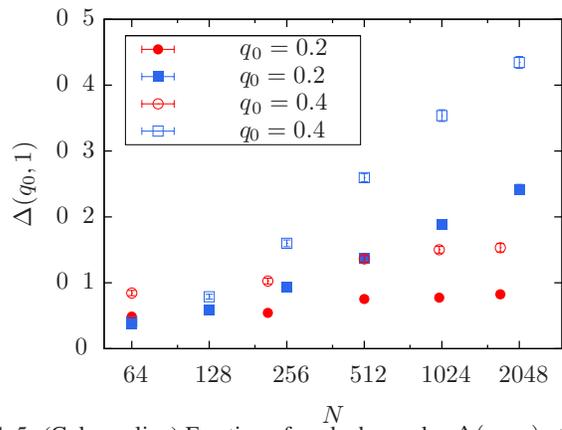}
\vspace*{-1.5em}
\caption{(Color online)
Fraction of peaked samples $\Delta(q_0,\kappa)$ at $T \approx 0.4T_c$ as
a function of $N$ for $\kappa=1$, $q_0=0.2$ and $0.4$.
\label{fig:frac}}
\end{figure}

Figure \ref{fig:frac} shows $\Delta(q_0,\kappa)$ as a function of system
size for  $q_0=0.2$ and $0.4$, as well as $\kappa = 1$
\cite{comment:errors}.  We found qualitatively similar results for other
values of $q_0$ and $\kappa$, as well as for lower temperatures. Our
most important observation is that the fraction of peaked samples
$\Delta(q_0,\kappa)$ is nearly constant and small for the EA model while
$\Delta(q_0,\kappa)$ is increasing over the same range of $N$ for the SK
model \cite{comment:aspel}. The result for the SK model is expected from
Parisi's RSB solution.  The contrasting result for the EA model suggests
that the number of pure states does not grow with the system size for
low $T$; a result consistent with the droplet and chaotic pairs
pictures.

\begin{figure}
\center
\includegraphics[width=0.95\columnwidth]{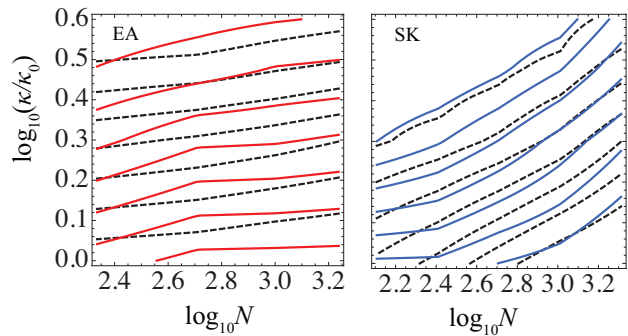}
\vspace*{-1.5em}
\caption{(Color online)
Contours of constant $\Delta$ for the EA model (left) and the SK model
(right) as a function of $\log_{10}(N)$ and $\log_{10}(\kappa/\kappa_0)$
with $\kappa_0=0.5$ and $1.5$ for $q_0=0.2$ and $1.0$, respectively.
The solid [dashed] lines are contours of constant $\Delta$ for $q_0=0.2$
[$q_0=1.0$] equally spaced in $\Delta$ \cite{comment:ranges}.
\label{fig:contour}}
\end{figure}

The difference in the behavior of $\Delta$ for the SK model in
comparison to the EA model might be explained by the fact that peaks
sharpen more quickly with $N$ for the SK than for the EA model (see
Fig.~\ref{fig:average} and Ref.~\cite{fernandez:12}. To study this
effect, we compare $\Delta$ for the two values, $q_0=0.2$ and $q_0=1$,
for each model separately.  For $q_0=1$, $\Delta$ is controlled by the
peaks at $\pm \qea$ and must converge to unity for both models because
for $N \rightarrow \infty$  the $\qea$ peaks become $\delta$ functions.
Figure \ref{fig:contour}, left [right] panel, shows contour plots of
constant $\Delta$ for the EA [SK] model.  The horizontal axis is the
logarithm of the number of spins and the vertical axis is the logarithm
of $\kappa/\kappa_0$ with $\kappa_0=0.5$ for $q_0=0.2$ and
$\kappa_0=1.5$ for $q_0=1$.  The curves are lines of constant $\Delta$
obtained from a linear interpolation of the data. Each set of curves are
equally spaced in $\Delta$ \cite{comment:ranges} with $\Delta$
decreasing as $\kappa$ increases. The dashed contours are for $q_0=1$
and thus include the $\qea$ peaks.  As expected, the dashed contours are
clearly increasing functions for {\em both} models although they rise
more rapidly for the SK model than for the EA model. The solid curves
are contours of constant $\Delta$ for $q_0=0.2$.  Close inspection of
the data reveals a {\em qualitative} difference between both models. For
large $N$ and large $\Delta$, the SK $q_0=0.2$ contours rise more
steeply than the corresponding $q_0=1$ contours, suggesting that not
only are peaks sharpening, but the number of peaks is also increasing.
In fact, Ref.~\cite{aspelmeier:08} shows that the number of peaks in
$P_J(q)$ should scale as $N^{1/6}$ for the SK model.  On the other hand,
for large $N$ and large $\Delta$, the EA contours for $q_0=0.2$ are
nearly flat, rising less steeply than for $q_0=1$, suggesting that the
number of peaks is either decreasing or staying constant.

\paragraph*{Conclusions.---} 
\label{sec:conclusions}

We introduce a statistic $\Delta$ that detects the fraction of samples
with $\delta$ function behavior in $P_{\cal J}(q)$ near the origin and
sharply distinguishes the RSB picture from scenarios with only a single
pair of states such as the droplet picture. While our results for the SK
model are consistent with RSB, as expected, the EA model does not
display a trend towards many pairs of pure states. These results lend
weight to the droplet and chaotic pairs pictures It is also possible
that for the EA model, $\Delta$ increases very slowly in $N$ and
ultimately converges to unity in agreement with the RSB picture.
However, our data show no indication of such trend.  It would be
interesting to perform a similar analysis with extremely large data sets
computed with special-purpose computers, such as the Janus machine
\cite{belletti:08}.

We thank J.~C.~Andresen, R.~S.~Andrist, D.~Fisher, D.~Huse, E.~Marinari,
M.~Moore, C.~Newman, G. Parisi, and D.~Stein for useful discussions.
H.G.K.~acknowledges support from the SNF (Grant No.~PP002-114713) and
the NSF (Grant No.~DMR-1151387). J.M.~and B.Y.~are supported in part by
the NSF (Grant No.~DMR-0907235).  We thank the Texas Advanced Computing
Center, ETH Zurich, and Texas A\&M University for providing HPC
resources.

\vspace*{-1.00em}

\bibliography{refs,comments}

\begin{thebibliography}{30}
\expandafter\ifx\csname natexlab\endcsname\relax\def\natexlab#1{#1}\fi
\expandafter\ifx\csname bibnamefont\endcsname\relax
  \def\bibnamefont#1{#1}\fi
\expandafter\ifx\csname bibfnamefont\endcsname\relax
  \def\bibfnamefont#1{#1}\fi
\expandafter\ifx\csname citenamefont\endcsname\relax
  \def\citenamefont#1{#1}\fi
\expandafter\ifx\csname url\endcsname\relax
  \def\url#1{\texttt{#1}}\fi
\expandafter\ifx\csname urlprefix\endcsname\relax\def\urlprefix{URL }\fi
\providecommand{\bibinfo}[2]{#2}
\providecommand{\eprint}[2][]{\url{#2}}

\bibitem[{\citenamefont{Binder and Young}(1986)}]{binder:86}
\bibinfo{author}{\bibfnamefont{K.}~\bibnamefont{Binder}} \bibnamefont{and}
  \bibinfo{author}{\bibfnamefont{A.~P.} \bibnamefont{Young}},
  \bibinfo{journal}{Rev. Mod. Phys.} \textbf{\bibinfo{volume}{58}},
  \bibinfo{pages}{801} (\bibinfo{year}{1986}).

\bibitem[{\citenamefont{Edwards and Anderson}(1975)}]{edwards:75}
\bibinfo{author}{\bibfnamefont{S.~F.} \bibnamefont{Edwards}} \bibnamefont{and}
  \bibinfo{author}{\bibfnamefont{P.~W.} \bibnamefont{Anderson}},
  \bibinfo{journal}{J. Phys. F: Met. Phys.} \textbf{\bibinfo{volume}{5}},
  \bibinfo{pages}{965} (\bibinfo{year}{1975}).

\bibitem[{\citenamefont{Sherrington and Kirkpatrick}(1975)}]{sherrington:75}
\bibinfo{author}{\bibfnamefont{D.}~\bibnamefont{Sherrington}} \bibnamefont{and}
  \bibinfo{author}{\bibfnamefont{S.}~\bibnamefont{Kirkpatrick}},
  \bibinfo{journal}{Phys. Rev. Lett.} \textbf{\bibinfo{volume}{35}},
  \bibinfo{pages}{1792} (\bibinfo{year}{1975}).

\bibitem[{\citenamefont{Parisi}(1979)}]{parisi:79}
\bibinfo{author}{\bibfnamefont{G.}~\bibnamefont{Parisi}},
  \bibinfo{journal}{Phys. Rev. Lett.} \textbf{\bibinfo{volume}{43}},
  \bibinfo{pages}{1754} (\bibinfo{year}{1979}).

\bibitem[{\citenamefont{Parisi}(1983)}]{parisi:83}
\bibinfo{author}{\bibfnamefont{G.}~\bibnamefont{Parisi}},
  \bibinfo{journal}{Phys. Rev. Lett.} \textbf{\bibinfo{volume}{50}},
  \bibinfo{pages}{1946} (\bibinfo{year}{1983}).

\bibitem[{\citenamefont{Katzgraber et~al.}(2006)\citenamefont{Katzgraber,
  K\"orner, and Young}}]{katzgraber:06}
\bibinfo{author}{\bibfnamefont{H.~G.} \bibnamefont{Katzgraber}},
  \bibinfo{author}{\bibfnamefont{M.}~\bibnamefont{K\"orner}}, \bibnamefont{and}
  \bibinfo{author}{\bibfnamefont{A.~P.} \bibnamefont{Young}},
  \bibinfo{journal}{Phys. Rev. B} \textbf{\bibinfo{volume}{73}},
  \bibinfo{pages}{224432} (\bibinfo{year}{2006}).

\bibitem[{\citenamefont{McMillan}(1984)}]{mcmillan:84b}
\bibinfo{author}{\bibfnamefont{W.~L.} \bibnamefont{McMillan}},
  \bibinfo{journal}{Phys. Rev. B} \textbf{\bibinfo{volume}{29}},
  \bibinfo{pages}{4026} (\bibinfo{year}{1984}).

\bibitem[{\citenamefont{Fisher and Huse}(1986)}]{fisher:86}
\bibinfo{author}{\bibfnamefont{D.~S.} \bibnamefont{Fisher}} \bibnamefont{and}
  \bibinfo{author}{\bibfnamefont{D.~A.} \bibnamefont{Huse}},
  \bibinfo{journal}{Phys. Rev. Lett.} \textbf{\bibinfo{volume}{56}},
  \bibinfo{pages}{1601} (\bibinfo{year}{1986}).

\bibitem[{\citenamefont{Fisher and Huse}(1987)}]{fisher:87}
\bibinfo{author}{\bibfnamefont{D.~S.} \bibnamefont{Fisher}} \bibnamefont{and}
  \bibinfo{author}{\bibfnamefont{D.~A.} \bibnamefont{Huse}},
  \bibinfo{journal}{J. Phys. A} \textbf{\bibinfo{volume}{20}},
  \bibinfo{pages}{L1005} (\bibinfo{year}{1987}).

\bibitem[{\citenamefont{Fisher and Huse}(1988)}]{fisher:88}
\bibinfo{author}{\bibfnamefont{D.~S.} \bibnamefont{Fisher}} \bibnamefont{and}
  \bibinfo{author}{\bibfnamefont{D.~A.} \bibnamefont{Huse}},
  \bibinfo{journal}{Phys. Rev. B} \textbf{\bibinfo{volume}{38}},
  \bibinfo{pages}{386} (\bibinfo{year}{1988}).

\bibitem[{\citenamefont{Bray and Moore}(1986)}]{bray:86}
\bibinfo{author}{\bibfnamefont{A.~J.} \bibnamefont{Bray}} \bibnamefont{and}
  \bibinfo{author}{\bibfnamefont{M.~A.} \bibnamefont{Moore}}, in
  \emph{\bibinfo{booktitle}{Heidelberg Colloquium on Glassy Dynamics and
  Optimization}}, edited by
  \bibinfo{editor}{\bibfnamefont{L.}~\bibnamefont{Van~Hemmen}}
  \bibnamefont{and}
  \bibinfo{editor}{\bibfnamefont{I.}~\bibnamefont{Morgenstern}}
  (\bibinfo{publisher}{Springer}, \bibinfo{address}{New York},
  \bibinfo{year}{1986}), p. \bibinfo{pages}{121}.

\bibitem[{\citenamefont{Newman and Stein}(1992)}]{newman:92}
\bibinfo{author}{\bibfnamefont{C.~M.} \bibnamefont{Newman}} \bibnamefont{and}
  \bibinfo{author}{\bibfnamefont{D.~L.} \bibnamefont{Stein}},
  \bibinfo{journal}{Phys. Rev. B} \textbf{\bibinfo{volume}{46}},
  \bibinfo{pages}{973} (\bibinfo{year}{1992}).

\bibitem[{\citenamefont{Newman and Stein}(1996)}]{newman:96}
\bibinfo{author}{\bibfnamefont{C.~M.} \bibnamefont{Newman}} \bibnamefont{and}
  \bibinfo{author}{\bibfnamefont{D.~L.} \bibnamefont{Stein}},
  \bibinfo{journal}{Phys. Rev. Lett.} \textbf{\bibinfo{volume}{76}},
  \bibinfo{pages}{515} (\bibinfo{year}{1996}).

\bibitem[{\citenamefont{Newman and Stein}(1998)}]{newman:98}
\bibinfo{author}{\bibfnamefont{C.~M.} \bibnamefont{Newman}} \bibnamefont{and}
  \bibinfo{author}{\bibfnamefont{D.~L.} \bibnamefont{Stein}},
  \bibinfo{journal}{Phys. Rev. E} \textbf{\bibinfo{volume}{57}},
  \bibinfo{pages}{1356} (\bibinfo{year}{1998}).

\bibitem[{\citenamefont{Katzgraber et~al.}(2001)\citenamefont{Katzgraber,
  Palassini, and Young}}]{katzgraber:01}
\bibinfo{author}{\bibfnamefont{H.~G.} \bibnamefont{Katzgraber}},
  \bibinfo{author}{\bibfnamefont{M.}~\bibnamefont{Palassini}},
  \bibnamefont{and} \bibinfo{author}{\bibfnamefont{A.~P.} \bibnamefont{Young}},
  \bibinfo{journal}{Phys. Rev. B} \textbf{\bibinfo{volume}{63}},
  \bibinfo{pages}{184422} (\bibinfo{year}{2001}).

\bibitem[{\citenamefont{Palassini and Young}(1999)}]{palassini:99}
\bibinfo{author}{\bibfnamefont{M.}~\bibnamefont{Palassini}} \bibnamefont{and}
  \bibinfo{author}{\bibfnamefont{A.~P.} \bibnamefont{Young}},
  \bibinfo{journal}{Phys. Rev. Lett.} \textbf{\bibinfo{volume}{83}},
  \bibinfo{pages}{5126} (\bibinfo{year}{1999}).

\bibitem[{\citenamefont{Krzakala and Martin}(2000)}]{krzakala:00}
\bibinfo{author}{\bibfnamefont{F.}~\bibnamefont{Krzakala}} \bibnamefont{and}
  \bibinfo{author}{\bibfnamefont{O.~C.} \bibnamefont{Martin}},
  \bibinfo{journal}{Phys. Rev. Lett.} \textbf{\bibinfo{volume}{85}},
  \bibinfo{pages}{3013} (\bibinfo{year}{2000}).

\bibitem[{\citenamefont{Aspelmeier et~al.}(2008)\citenamefont{Aspelmeier,
  Billoire, Marinari, and Moore}}]{aspelmeier:08}
\bibinfo{author}{\bibfnamefont{T.}~\bibnamefont{Aspelmeier}},
  \bibinfo{author}{\bibfnamefont{A.}~\bibnamefont{Billoire}},
  \bibinfo{author}{\bibfnamefont{E.}~\bibnamefont{Marinari}}, \bibnamefont{and}
  \bibinfo{author}{\bibfnamefont{M.~A.} \bibnamefont{Moore}},
  \bibinfo{journal}{J. Phys. A: Math. Theor.} \textbf{\bibinfo{volume}{41}},
  \bibinfo{pages}{324008} (\bibinfo{year}{2008}).

\bibitem[{\citenamefont{Hukushima and Nemoto}(1996)}]{hukushima:96}
\bibinfo{author}{\bibfnamefont{K.}~\bibnamefont{Hukushima}} \bibnamefont{and}
  \bibinfo{author}{\bibfnamefont{K.}~\bibnamefont{Nemoto}},
  \bibinfo{journal}{J. Phys. Soc. Jpn.} \textbf{\bibinfo{volume}{65}},
  \bibinfo{pages}{1604} (\bibinfo{year}{1996}).

\bibitem[{\citenamefont{Hukushima et~al.}(1998)\citenamefont{Hukushima,
  Takayama, and Yoshino}}]{hukushima:98}
\bibinfo{author}{\bibfnamefont{K.}~\bibnamefont{Hukushima}},
  \bibinfo{author}{\bibfnamefont{H.}~\bibnamefont{Takayama}}, \bibnamefont{and}
  \bibinfo{author}{\bibfnamefont{H.}~\bibnamefont{Yoshino}},
  \bibinfo{journal}{J. Phys. Soc. Jpn.} \textbf{\bibinfo{volume}{67}},
  \bibinfo{pages}{12} (\bibinfo{year}{1998}).

\bibitem[{com({\natexlab{a}})}]{comment:temps}
\bibinfo{note}{Tests at low $T$ give qualitatively similar results.}

\bibitem[{com({\natexlab{b}})}]{comment:pq}
\bibinfo{note}{Because $P(q)$ is an even function, we show one side of $P(q)$
  for each model.}

\bibitem[{\citenamefont{Alvarez Ba\~nos et~al.}(2010)\citenamefont{Alvarez
  Ba\~nos, Cruz, Fernandez, Gordillo-Guerrero, Gil-Narvion, Guidetti, Maiorano,
  Mantovani, Marinari, Martin-Mayor et~al.}}]{alvarez:10}
\bibinfo{author}{\bibfnamefont{R.}~\bibnamefont{Alvarez Ba\~nos}},
  \bibinfo{author}{\bibfnamefont{A.}~\bibnamefont{Cruz}},
  \bibinfo{author}{\bibfnamefont{L.~A.} \bibnamefont{Fernandez}},
  \bibinfo{author}{\bibfnamefont{A.}~\bibnamefont{Gordillo-Guerrero}},
  \bibinfo{author}{\bibfnamefont{J.~M.} \bibnamefont{Gil-Narvion}},
  \bibinfo{author}{\bibfnamefont{M.}~\bibnamefont{Guidetti}},
  \bibinfo{author}{\bibfnamefont{A.}~\bibnamefont{Maiorano}},
  \bibinfo{author}{\bibfnamefont{F.}~\bibnamefont{Mantovani}},
  \bibinfo{author}{\bibfnamefont{E.}~\bibnamefont{Marinari}},
  \bibinfo{author}{\bibfnamefont{V.}~\bibnamefont{Martin-Mayor}},
  \bibnamefont{et~al.}, \bibinfo{journal}{J. Stat. Mech.}
  \textbf{\bibinfo{volume}{\normalfont{P05002}}} (\bibinfo{year}{2010}).

\bibitem[{\citenamefont{Hartmann}(1999)}]{hartmann:99}
\bibinfo{author}{\bibfnamefont{A.~K.} \bibnamefont{Hartmann}},
  \bibinfo{journal}{Phys. Rev. E} \textbf{\bibinfo{volume}{59}},
  \bibinfo{pages}{84} (\bibinfo{year}{1999}).

\bibitem[{com({\natexlab{c}})}]{comment:errors}
\bibinfo{note}{The error bars in Fig.~\ref{fig:frac} are one standard deviation
  statistical errors due to the finite number of samples. There are also errors
  due to the finite length of the data collection. To estimate these errors we
  measured $\Delta^+(q_0,\kappa)$ and $\Delta^-(q_0,\kappa)$, defined as in
  Eq.~(\ref{eq:ff}) but from the $q>0$ and $q<0$ components of $P_{\cal J}(q)$,
  respectively. These are expected to be reasonably independent and their
  differences provide an estimate of the error due to finite run lengths. For
  all sizes, the average absolute difference between these quantities,
  $(|\Delta^+(q_0,\kappa)-\Delta(q_0,\kappa)|
  +|\Delta^-(q_0,\kappa)-\Delta(q_0,\kappa)|)/2$ is less than the statistical
  error.}

\bibitem[{com({\natexlab{d}})}]{comment:aspel}
\bibinfo{note}{Similar results for the SK model were obtained in
  Ref.~\cite{aspelmeier:08-ea} by counting individual peaks.}

\bibitem[{com({\natexlab{e}})}]{comment:ranges}
\bibinfo{note}{For the EA model the ranges are $(0.2,0.8)$ for $q_0=1$ and
  $(0.02,0.14)$ for $q_0=0.2$. For the SK model the ranges are $(0.05,0.77)$
  for $q_0=1$ and $(0.06,0.288)$ for $q_0=0.2$.}

\bibitem[{\citenamefont{{Fern{\'a}ndez} and {Alonso}}(2012)}]{fernandez:12}
\bibinfo{author}{\bibfnamefont{J.~F.} \bibnamefont{{Fern{\'a}ndez}}}
  \bibnamefont{and} \bibinfo{author}{\bibfnamefont{J.~J.}
  \bibnamefont{{Alonso}}} (\bibinfo{year}{2012}),
  \bibinfo{note}{(arXiv:cond-mat/1207.4008)}.

\bibitem[{\citenamefont{{Belletti} et~al.}(2008)\citenamefont{{Belletti},
  {Cotallo}, {Cruz}, {Fern{\'a}ndez}, {Gordillo}, {Maiorano}, {Mantovani},
  {Marinari}, {Mart{\'{\i}}n-Mayor}, {Mu{\~n}oz-Sudupe} et~al.}}]{belletti:08}
\bibinfo{author}{\bibfnamefont{F.}~\bibnamefont{{Belletti}}},
  \bibinfo{author}{\bibfnamefont{M.}~\bibnamefont{{Cotallo}}},
  \bibinfo{author}{\bibfnamefont{A.}~\bibnamefont{{Cruz}}},
  \bibinfo{author}{\bibfnamefont{L.~A.} \bibnamefont{{Fern{\'a}ndez}}},
  \bibinfo{author}{\bibfnamefont{A.}~\bibnamefont{{Gordillo}}},
  \bibinfo{author}{\bibfnamefont{A.}~\bibnamefont{{Maiorano}}},
  \bibinfo{author}{\bibfnamefont{F.}~\bibnamefont{{Mantovani}}},
  \bibinfo{author}{\bibfnamefont{E.}~\bibnamefont{{Marinari}}},
  \bibinfo{author}{\bibfnamefont{V.}~\bibnamefont{{Mart{\'{\i}}n-Mayor}}},
  \bibinfo{author}{\bibfnamefont{A.}~\bibnamefont{{Mu{\~n}oz-Sudupe}}},
  \bibnamefont{et~al.}, \bibinfo{journal}{Comp. Phys. Comm.}
  \textbf{\bibinfo{volume}{178}}, \bibinfo{pages}{208} (\bibinfo{year}{2008}).

\bibitem[{\citenamefont{Aspelmeier~{\em et al.}}(2008)}]{aspelmeier:08-ea}
\bibinfo{author}{\bibfnamefont{T.}~\bibnamefont{Aspelmeier~{\em et al.}}},
  \bibinfo{journal}{J. Phys. A: Math. Theor.} \textbf{\bibinfo{volume}{41}},
  \bibinfo{pages}{324008} (\bibinfo{year}{2008}).

\end{thebibliography}

\end{document}